\documentclass{article}

\PassOptionsToPackage{numbers, compress}{natbib}

 \usepackage[preprint]{nips_2018}

\usepackage[utf8]{inputenc} 
\usepackage[T1]{fontenc}    
\usepackage{hyperref}       
\usepackage{url}            
\usepackage{booktabs}       
\usepackage{amsfonts}       
\usepackage{nicefrac}       
\usepackage{microtype}      
\usepackage{hhline}

\usepackage{amsmath}
\usepackage{amsfonts}
\usepackage{amssymb}
\usepackage{bm}
\usepackage{graphicx}
\usepackage{amsthm}

\usepackage{tikz}
\usetikzlibrary{decorations.markings}
\usetikzlibrary{patterns}
\usepackage{pgfplots}
\pgfplotsset{compat=1.14}
\usepackage{multirow}
\usepackage{varwidth}
\usetikzlibrary{arrows}
\usetikzlibrary{shapes.geometric}
\newcommand\msheight{3}

\tikzset{lstm_cell/.style={draw, minimum size=2em, pattern=north west lines, pattern color=gray, rounded corners}}
\tikzset{mass_spec/.style={draw, rectangle, minimum height=4em, minimum width=12em}}
\tikzset{mass_spec_reading/.style={color=red, line width=1.6}}
\tikzset{mass_spec_concatenation/.style={color=blue, line width=1.6}}
\tikzset{cross_correlator/.style={draw, rectangle, minimum height=0.5em, minimum width=13em, pattern=north west lines, pattern color=gray}}
\tikzset{readout/.style={draw, trapezium, minimum height=0.5em, minimum width=14.2em, trapezium left angle=10, trapezium right angle=10, pattern color=gray, pattern=north west lines}}
\tikzset{score/.style={draw, rectangle, minimum height=0.5em, minimum width=0.5em, fill=black!10!green, rounded corners}}

\title{Peptide-Spectra Matching from Weak Supervision}

\author{
Samuel S. Schoenholz\\
Google Brain\\
\texttt{schsam@google.com}
\And
Sean Hackett\\
Calico Life Sciences LLC\\
\texttt{sean@calicolabs.com}
\And
Laura Deming\\
Longevity Fund\\
\texttt{ldeming.www@gmail.com}
\And
Eugene Melamud\\
Calico Life Sciences LLC\\
\texttt{eugene@calicolabs.com}
\And
Navdeep Jaitly\\
Google Brain\\
\texttt{ndjaitly@google.com}
\And
Fiona McAllister\\
Calico Life Sciences LLC\\
\texttt{fiona@calicolabs.com}
\And
Jonathon O'Brien\\
Calico Life Sciences LLC\\
\texttt{obrienj@calicolabs.com}
\And
George Dahl\\
Google Brain\\
\texttt{gdahl@google.com}
\And
Bryson Bennett\\
Calico Life Sciences LLC\\
\texttt{bryson@calicolabs.com}
\And
Andrew M. Dai\\
Google Brain\\
\texttt{adai@google.com}
\And
Daphne Koller\\
Insitro\thanks{Work done while at Calico Life Sciences LLC.}\\
\texttt{daphne@insitro.com}
}

\begin{document}

\maketitle

\begin{abstract}
As in many other scientific domains, we face a fundamental problem when using machine learning to identify proteins from mass spectrometry data: large ground truth datasets mapping inputs to correct outputs are extremely difficult to obtain. Instead, we have access to imperfect hand-coded models crafted by domain experts. In this paper, we apply deep neural networks to an important step of the protein identification problem, the pairing of mass spectra with short sequences of amino acids called peptides. We train our model to differentiate between top scoring results from a state-of-the art classical system and hard-negative second and third place results. Our resulting model is much better at identifying peptides with spectra than the model used to generate its training data. In particular, we achieve a 43\% improvement over standard matching methods and a 10\% improvement over a combination of the matching method and an industry standard cross-spectra reranking tool. Importantly, in a more difficult experimental regime that reflects current challenges facing biologists, our advantage over the previous state-of-the-art grows to 15\% even after reranking. We believe this approach will generalize to other challenging scientific problems.
\end{abstract}
\section{Introduction}

Historically, analysis of complex scientific data has been made possible by domain experts using theory and expert knowledge to create simple causal models that can be used to make inferences from measurements. In these cases, there is often no ground truth associated with the data and a great deal of effort goes into quantifying the performance of these hand-made models. Examples of this situation are widespread but include: scattering data in high-energy physics experiments~\citep{aad2015combined}, measurements from gravitational wave observatories~\citep{abbott2016observation}, and cryo-electron microscopy~\citep{henderson1975three}. The difficulty of quantitative evaluation complicates the use of machine learning at the cutting-edge of science.

Simultaneously, machine learning models based on deep neural networks have achieved impressive performance across a variety of tasks in the natural sciences in cases where ground truth is available. Some examples of these recent successes include predicting the properties of small molecules~\citep{gilmer2017neural}, predicting the properties of drugs~\citep{schneider2017} and materials~\citep{sendek2017holistic}, and predicting the local structure of proteins~\citep{secstruct}. Ideally we would be able to achieve similar success even when using weaker supervision. In this work, we tackle this more general problem of weak supervision in the context of a scientific problem of practical importance: identifying proteins in biological samples from mass spectrometry measurements. Our approach is to leverage noisy labels from classical database matching algorithms to train a machine learning algorithm whose performance is \emph{significantly} better than the tool used to construct the labels.

Answering fundamental questions in biology such as how cellular processes are regulated or how these processes are altered in disease requires an accurate picture of an organism's biochemical state. Since most cellular processes of interest are implemented by proteins, the state of proteins in a cell is extremely relevant information if we want to understand how cells operate. Cells, however, are complex environments that contain large numbers of proteins. The inverse problem of identifying which proteins are present in a given cell is complex and has received a significant amount of attention. A crucial step in modern workflows involves associating short sequences of amino acids, called peptides, with readings from a mass spectrometer (which essentially produces a histogram of the different ways that the peptide can fragment). If these peptides can successfully be identified, then other models can reconstruct the protein population. Unfortunately, ground truth pairings between peptides and spectra are not readily obtained. Therefore, if machine learning is to be successful in this domain we must be able to use existing noisy labels to make progress.

The primary contributions of this manuscript are as follows. We first introduce a method for training general machine learning models to identify peptides corresponding to a given mass spectrum based on noisy labels from classical algorithms. We include a detailed discussion of pitfalls that can occur when attempting to apply deep learning methods in this framework along with solutions to avoid various modes of failure. Next, we introduce a family of models that we believe have a strong inductive bias towards the problem of matching sequences with histogram readouts. Finally, we discuss the performance of our model, show that it significantly outperforms classical approaches, and demonstrate that this performance gap widens as the problem becomes harder.

\section{Related Work}
Our paper shows how unlabeled mass spectral data can be assigned peptide labels using prior models and how new models can be trained from the generated labels. Such \emph{weak},  \emph{pseudo} and \emph{self} labeling strategies are not new to the area of machine learning \citep{pseudolabels,liao2013large,mcclosky2006effective}. For example, \citet{liao2013large} applies an existing model to YouTube videos to generate possible transcripts and filters out matches that are presumed to be bad. Good matches are used to train a subsequent model, producing improvements over the previous model. Pseudo-labeling has been used previously in mass spectrometry for retention time prediction with a neural network, where a dataset of peptide sequence, retention time pairs was generated from large datasets using stringent rules applied to SEQUEST results \citep{petritis2006improved}. Percolator, an algorithm for peptide identification from mass spectrometry, also used a \emph{pseudo} labelling strategy to train their model \citep{Kall:2008is}. However, Percolator uses a fixed set of fairly high level features of the matches in their support vector machine, while this work uses deep learning strategies to discover features and scoring functions directly from the data -- the mass spectrum itself. Additionally, a similar labeling strategy to that proposed here has been used in physics to identify defects in glasses from noisy measurements of relaxation events~\citep{schoenholz2016}.

The use of deep learning models for mapping sequences to structured outputs is also well studied\citep{sutskever2014sequence,vinyals2015grammar,oord2016wavenet,secstruct}. Typically, however, such models are applied in a blackbox manner; in contrast, in this paper, we introduce an architecture that is partially a black-box, and is partially manually designed to reflect the structure of the peptide fragment problem --- namely that the fragmentation pattern of a peptide is a sum of the fragmentation pattern of the N- and C-terminus subsequences of the peptide. However, our model is modular, and can be easily extended to other input structures, such as graphs (or chemical structures) using graph embeddings \citep{graphNN} and as a result can be applied to other mass spectrometry modalities such as metabolomics and scanning electron microscopy. 

\citet{Tran:2017id} have previously applied deep learning to peptide-spectra matching. However, they focused on the \textit{de novo} sequencing problem where the peptide search space is unconstrained and is consequently a substantially more difficult problem. While they demonstrated a substantial improvement over the state-of-the-art, results on the \textit{de novo} sequencing remain too poor to be used in practice.

As an additional note, our method of comparing best match sequences to second and third-best sequence matches for a spectral output, might also be appropriate for hot-word matching models in speech recognition, where fixed length audio snippets are often compared to a set of candidate hot-words \citep{bengio2014word}. Another possible application of our model that identifies structured inputs with a distribution might be the association of crystal structures with data from scattering experiments~\citep{ashcroft1978solid}.

\subsection{Experimental Proteomics}
\begin{figure}
\centering
\includegraphics[width=1.0\textwidth]{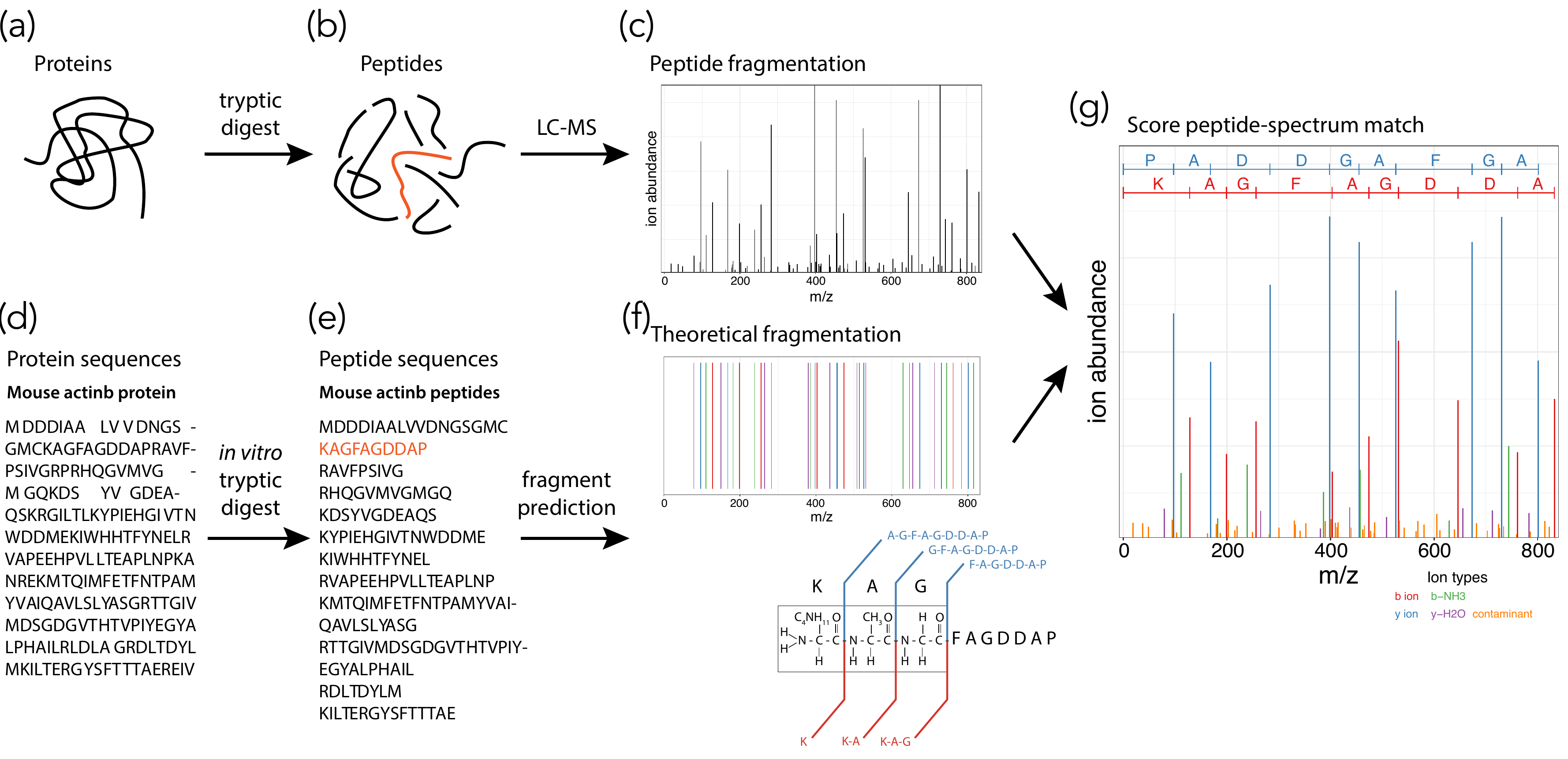}
\caption{\textbf{Overview of fragmentation matching in LC-MS proteomics.} Experimental fragmentation spectra of unknown proteins (a) are generated by isolating individual peptides (b) and then fragmenting the peptide into pieces in a mass spectrometer (c). To narrow down the number of feasible candidates for a fragmentation event, the genome of the organism of interest (d) can be digested to yield a set of feasible peptides (e). Each peptide's common fragments can be predicted based on its sequence, since  peptides primarily fragment along the C-N bonds in the backbone and often lose either water or ammonia in the process (f). During the peptide-spectrum matching process an experimental fragmentation spectra is compared to the theoretical spectra of a candidate peptide to determine the most likely match (g).}
%
%
\label{fig:frag_overview}
\end{figure}

In proteomics, the goal is to identify protein populations occurring in organisms. Here we give a brief summary of experimental proteomics. In the supplementary information we include a slightly more extensive description. Figure~\ref{fig:frag_overview} shows a simplified schematic of a common pipeline for protein identification used in proteomics called liquid chromatography-mass spectrometry (LC-MS).
Proteins from complex biological samples are isolated with biochemical techniques, and separated into less complex mixtures based on their properties (Figure~\ref{fig:frag_overview}a). These mixtures of proteins are digested into pieces called peptides (\ref{fig:frag_overview}b) and further separated by liquid chromatography based on their biochemical properties. The peptides are individually (hopefully) injected into a mass spectrometer and fragmented into smaller pieces. The mass to charge ratio of these fragments is then measured to produce a histogram of the relative intensities of the fragments (\ref{fig:frag_overview}c). The goal of a modern LC-MS pipeline is then to infer the original protein population from these histograms. 

\section{Methods}

\subsection{DeepMatch: A model for peptide-spectra matching}


In this section, we will describe our model which we refer to as DeepMatch. Suppose we have scans from a mass spectrometer that we wish to associate with peptides. Generally a scan comes as a set of pairs $(m_i, I_i)$ where $m_i$ is the $m/z$ (mass-to-charge ratio) of a peak in the spectrum and $I_i$ is the measured intensity of the peak. We will discretize this measurement into $M$ different $m/z$ bins of width $\Delta M$ so that $M_i = M_{\text{min}} + i\Delta M$ are the bin edges; thus the reading from the mass spectrometer will be represented by a vector $\hat y\in\mathbb R^M_{\ge 0}$ such that $\hat y_i$ is the summed intensity at each bin. In these experiments $M = 3800$, $M_{\text{min}} = 100$ Daltons, and $\Delta M = 0.5$ Daltons. We think of the mass spectrometer as measuring the distribution over all the different ways in which the peptide can fragment and the different $m/z$ ratios that the fragments can have (either due to variable charges, neutral losses, etc). To this end we turn the intensity measurements into a probability distribution by normalizing the reading to give $\hat p_i = \hat y_i / ||\hat y||_1$.

Simultaneously we can represent the peptide as a sequence of $L$ (possibly modified) amino acids, $x_1\cdots x_L$. We will typically use a one-hot encoding of the amino acids that we will augment with additional side information such as the mass of the peptide, the precursor charge, the retention time, and the hydrophobicity. Thus, we embed each amino acid $x_a\in\mathbb R^A$.

To train our model, we will construct a dataset of sequence -- distribution pairs. Some of these pairs will be real associations and have an assigned label $t = 1$. Other pairs will be fake and have a label $t = 0$. Our goal will be to train a model that can distinguish between the real and fake examples.

We propose a model family with four distinct pieces that we discuss schematically here. In the supplementary information we discuss several different components that we explored for each piece of the network. Note, that different components can be swapped out easily as long as their input / output dimensions agree.

\emph{Fragment Representation}: A function $F:\mathbb R^{L\times A}\to \mathbb R^{(L+1)\times F}$, maps sequences of amino acids to a ``fragment'' representation, of dimension $F$, that we hope will contain information about the probabilities of different fragmentations as well as their $m/z$ values. We will write $(f_1, \cdots, f_{L+1}) = F(x_1,\cdots, x_L)$. Typically we will take $F$ to be a bidirectional-LSTM~\cite{hochreiter1997}. If we wished to try to apply these methods to molecules that could not naturally be expressed as a sequence we could also have $F$ be an MPNN~\cite{graphNN,gilmer2017neural}.

\emph{Spectral Representation}: The second piece of our model is a composition of a fully connected neural network, $Y:\mathbb R^{(L+1)\times F}\to\mathbb R^{M\times H}$, which maps the fragment representation to a ``spectral'' representation with embedding dimension $H$. This spectral representation associates the different fragments with $m/z$ bins. Generally we will write $(y_1,\cdots, y_M) = Y(f_1,\cdots, f_{L+1})$. See the supplementary information for more details.

\emph{Readout}: Next, our model will compare the fragment representation of the peptide, $y_i$, with the measured spectrum $\hat y_i$ to arrive at a score, $s\in\mathbb R$ using a score function $S:\mathbb R^{M\times (H+1)}\to\mathbb R$. We will typically write $s = S(\hat y, y)$. We will typically use the normalized version of the spectrum, $\hat p_i$ instead of the raw intensities. In practice we will let $S$ be based on the VGG network architecture~\cite{simonyan2014very} similar to those found in image recognition systems.

\emph{Loss}: Finally, we will compute a per-example loss that we hope to extremize using our score. We will take this to be a cross entropy loss. To define the cross entropy loss we construct a probability of a match being real using the score as a logit, $\hat p = \sigma(s)$. We can then define the cross-entropy, $L = -\sum_i [p_i\log\hat p_i + (1-p_i)\log(1-\hat p_i)]$ where $p_i$ can be the label, $t_i$. However, we will often have information about how likely a given sequence-distribution pair is of being real. In that case $p_i$ can be set to this likelihood to give a form of label smoothing that we have found to be helpful in practice.

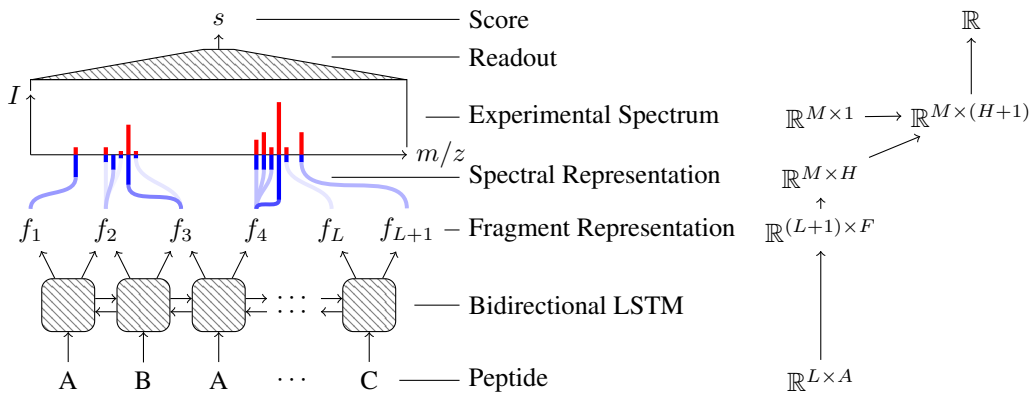
\begin{figure}
\begin{tikzpicture}

\node (A) at (0, 0) {A};
\node (B) at (1, 0) {B};
\node (C) at (2, 0) {A};
\node (dots) at (3, 0) {$\cdots$};
\node (D) at (4, 0) {C};


\node[lstm_cell] (A_lstm) at (0, 1) {};
\node[lstm_cell] (B_lstm) at (1, 1) {};
\node[lstm_cell] (C_lstm) at (2, 1) {};
\node[lstm_cell] (D_lstm) at (4, 1) {};


\node[transform canvas={yshift=0.25em}] (CD_dots_top) at (3, 1) {$\cdots$};
\node[transform canvas={yshift=-0.25em}] (CD_dots_bottom) at (3, 1) {$\cdots$};


\draw[->] (A) -- (A_lstm);
\draw[->] (B) -- (B_lstm);
\draw[->] (C) -- (C_lstm);
\draw[->] (D) -- (D_lstm);


\draw[->, transform canvas={yshift=0.25em}] (A_lstm) -- (B_lstm);
\draw[->, transform canvas={yshift=-0.25em}] (B_lstm) -- (A_lstm);
\draw[->, transform canvas={yshift=0.25em}] (B_lstm) -- (C_lstm);
\draw[->, transform canvas={yshift=-0.25em}] (C_lstm) -- (B_lstm);

\draw[->, transform canvas={yshift=0.25em}] (C_lstm) -- (CD_dots_top);
\draw[->, transform canvas={yshift=-0.25em}] (CD_dots_top) -- (C_lstm);

\draw[->, transform canvas={yshift=-0.25em}] (D_lstm) -- (CD_dots_top);
\draw[->, transform canvas={yshift=0.25em}] (CD_dots_top) -- (D_lstm);


\node (z_0) at (-0.5, 2) {$f_1$};
\node (z_1) at (0.5, 2) {$f_2$};
\node (z_2) at (1.5, 2) {$f_3$};
\node (z_3) at (2.5, 2) {$f_4$};
\node (z_4) at (3.5, 2) {$f_L$};
\node (z_5) at (4.5, 2) {$f_{L+1}$};


\draw[->] (A_lstm) -- (z_0);
\draw[->] (A_lstm) -- (z_1);
\draw[->] (B_lstm) -- (z_1);
\draw[->] (B_lstm) -- (z_2);
\draw[->] (C_lstm) -- (z_2);
\draw[->] (C_lstm) -- (z_3);
\draw[->] (D_lstm) -- (z_4);
\draw[->] (D_lstm) -- (z_5);

\draw[->] (-0.5, \msheight) -- (-0.5, \msheight+0.8) node[left] {$I$};
\draw[->] (-0.5, \msheight) -- (4.5, \msheight) node[right] {$m/z$};

\draw[mass_spec_reading] (0.1, \msheight) -- (0.1, \msheight + 0.1);

\draw[mass_spec_reading] (0.5, \msheight) -- (0.5, \msheight + 0.1);
\draw[mass_spec_reading] (0.7, \msheight) -- (0.7, \msheight + 0.05);
\draw[mass_spec_reading] (0.8, \msheight) -- (0.8, \msheight + 0.4);
\draw[mass_spec_reading] (0.9, \msheight) -- (0.9, \msheight + 0.05);

\draw[mass_spec_reading] (2.5, \msheight) -- (2.5, \msheight + 0.2);
\draw[mass_spec_reading] (2.6, \msheight) -- (2.6, \msheight + 0.3);
\draw[mass_spec_reading] (2.7, \msheight) -- (2.7, \msheight + 0.1);
\draw[mass_spec_reading] (2.8, \msheight) -- (2.8, \msheight + 0.7);
\draw[mass_spec_reading] (2.9, \msheight) -- (2.9, \msheight + 0.1);

\draw[mass_spec_reading] (3.1, \msheight) -- (3.1, \msheight + 0.3);

\draw[mass_spec_concatenation] (0.1, \msheight) -- (0.1, \msheight - 0.3);

\draw[mass_spec_concatenation] (0.5, \msheight) -- (0.5, \msheight - 0.1);
\draw[mass_spec_concatenation] (0.6, \msheight) -- (0.6, \msheight - 0.2);
\draw[mass_spec_concatenation] (0.7, \msheight) -- (0.7, \msheight - 0.05);
\draw[mass_spec_concatenation] (0.8, \msheight) -- (0.8, \msheight - 0.4);
\draw[mass_spec_concatenation] (0.9, \msheight) -- (0.9, \msheight - 0.05);

\draw[mass_spec_concatenation] (2.5, \msheight) -- (2.5, \msheight - 0.2);
\draw[mass_spec_concatenation] (2.6, \msheight) -- (2.6, \msheight - 0.2);
\draw[mass_spec_concatenation] (2.7, \msheight) -- (2.7, \msheight - 0.2);
\draw[mass_spec_concatenation] (2.8, \msheight) -- (2.8, \msheight - 0.6);
\draw[mass_spec_concatenation] (2.9, \msheight) -- (2.9, \msheight - 0.1);

\draw[mass_spec_concatenation] (3.1, \msheight) -- (3.1, \msheight - 0.1);


\draw[line width=1.5, color=blue!40] (z_0) to[out=90, in=-90] (0.1, \msheight - 0.3);

\draw[line width=1.5, color=blue!10] (z_4) to[out=90, in=-90] (2.9, \msheight - 0.1);

\draw[line width=1.5, color=blue!5] (z_1) to[out=90, in=-90] (0.7, \msheight - 0.05);
\draw[line width=1.5, color=blue!10] (z_1) to[out=90, in=-90] (0.5, \msheight - 0.1);
\draw[line width=1.5, color=blue!25] (z_1) to[out=90, in=-90] (0.6, \msheight - 0.2);

\draw[line width=1.5, color=blue!10] (z_2) to[out=90, in=-90] (0.9, \msheight - 0.05);
\draw[line width=1.5, color=blue!50] (z_2) to[out=90, in=-90] (0.8, \msheight - 0.4);

\draw[line width=1.5, color=blue!25] (z_3) to[out=90, in=-90] (2.5, \msheight - 0.2);
\draw[line width=1.5, color=blue!25] (z_3) to[out=90, in=-90] (2.7, \msheight - 0.2);
\draw[line width=1.5, color=blue!25] (z_3) to[out=90, in=-90] (2.6, \msheight - 0.2);
\draw[line width=1.5, color=blue!70] (z_3) to[out=90, in=-90] (2.8, \msheight - 0.6);

\draw[line width=1.5, color=blue!30] (z_5) to[out=90, in=-90] (3.1, \msheight - 0.1);



\draw (-0.5, \msheight + 0.85) to (-0.5, \msheight + 1.0) to (2, \msheight + 1.0);
\draw (4.5, \msheight + 0.1) to (4.5, \msheight + 1.0) to (2, \msheight + 1.0);

\node[readout] (readout) at (2, \msheight + 1.2) {};

\node (score) at (2, \msheight + 1.8) {$s$};

\draw[->] (readout) -- (score);

\draw (4.4, 0.0) -- (5.2, 0.0) node[right] {Peptide};

\draw (4.6, 1.0) -- (5.2, 1.0) node[right] {Bidirectional LSTM};

\draw (5.0, 2.0) -- (5.2, 2.0) node[right] {\begin{varwidth}{12em}Fragment Representation \end{varwidth}};

\draw (3.5, 2.7) -- (5.2, 2.7) node[right] {\begin{varwidth}{12em}Spectral Representation \end{varwidth}};

\draw (4.75,  \msheight + 0.5) -- (5.2,  \msheight + 0.5) node[right] {Experimental Spectrum};

\draw (3.5,  \msheight + 1.3) -- (5.2,  \msheight + 1.3) node[right] {Readout};

\draw (2.5,  \msheight + 1.8) -- (5.2,  \msheight + 1.8) node[right] {Score};

\node (peptide_dimension) at (10, 0) {$\mathbb R^{L\times A}$};
\node (fragment_dimension) at (10, 2) {$\mathbb R^{(L+1)\times F}$};
\node (theoretical_spectra_dimension) at (10, 2.7) {$\mathbb R^{M\times H}$}; 
\node (experimental_spectra_dimension) at (10, \msheight + 0.5) {$\mathbb R^{M\times 1}$}; 
\node (spectra_dimension) at (12, \msheight*0.5 + 0.75*0.5 + 3.3*0.5) {$\mathbb R^{M\times (H+1)}$}; 
\node (score_dimension) at (12, \msheight + 1.8) {$\mathbb R$}; 

\draw[->] (peptide_dimension) -- (fragment_dimension);
\draw[->] (fragment_dimension) -- (theoretical_spectra_dimension);
\draw[->] (theoretical_spectra_dimension) -- (spectra_dimension);
\draw[->] (experimental_spectra_dimension) -- (spectra_dimension);
\draw[->] (spectra_dimension) -- (score_dimension);
\end{tikzpicture}
\centering
\caption{\textbf{Components of DeepMatch along with shapes.}}
\label{fig:model_diagram}
\end{figure}
We can see a diagram of this class of models in Figure~\ref{fig:model_diagram}. We will go into details on the specific choices of these components for the problem of peptide-spectrum identification in the supplementary information.

\subsection{Weak Supervision}
As discussed above, a fundamental issue in the application of deep learning to problems across a variety of domains is the lack of ground truth labels. Here, we offer a discussion of this problem that ought to be general enough to be applicable in a wide range of situations. 

Suppose we have a generative process so that a distribution, $p(y|x)$, is generated by a latent variable $x$. In the current setting, $x$ will be a sequence of amino acids and $p(y|x)$ will be the mass spectrum that results from fragmenting that sequence. Further, let us assume that we have access to a noisy labeling function $f$ that assigns scores to sequence-distribution pairs. We take the noisy labeling function to be well behaved in the sense that if $ f(x, q(y)) > f(x, q'(y))$ then $x$ is more likely to have generated $q(y)$ than $q'(y)$.

For a given empirical distribution $q(y)$, we can therefore identify the most likely sequence to have produced $q(y)$ by extremizing, $\hat x = \text{argmax}_x f(x, q(y))$. This, in turn gives an associated score for the pair, $\hat s = f(\hat x, q(y))$. Thus, for any dataset of unlabeled distributions we can construct a labeled dataset where we additionally pair each example with an associated score. To construct our training set, we then choose a threshold $s^*$ and pick only those elements of the dataset whose score is greater than $s^*$. By increasing the threshold, this allows us to tune between dataset size and match confidence. In cases where the score can be explicitly associated with a probability of a correct match, we can use this probability to perform label-smoothing as discussed above.

\section{Methods}
\subsection{Peptide Spectrum Matching}
In order to make use of the quantitative information present in LC-MS proteomics data, we need to know which peptides, and by extension which proteins, were detected in a sample. This occurs by predicting which peptide generated each of the many fragmentation spectra produced in an experiment. Fragmentation spectra are informative of peptide sequence because peptides fragment in a probabilistic way, dependent on their sequence. For peptides that fragment cleanly, and produce a high-quality spectra, the masses of individual amino acids in the sequence can essentially be read-off from the spacing between prominent neighbouring peaks. However, most peptides do not fragment or ionize equally well at each bond; this can often be a function of their sequence (such as electron-affinity of subsequences). Thus some fragments will be missing in spectra. Further, contamination from other peptides of the same parent $m/z$ can produce noisy peaks, making the matching of fragmentation spectra to peptide sequences an ongoing challenge. While knowledge of simple fragmentation patterns is well known, one of our primary goals from this paper is to learn discriminative models of fragmentation patterns that improve identification, using recurrent sequence models such as LSTMs. Here, we offer a brief account of current techniques for associating peptides with spectra with reference to Figure~\ref{fig:frag_overview}. However, see~\citet{Steen:2004eu} for a more detailed account.

To solve the matching problem, classical approaches first construct a list of possible proteins present in the sample (Figure~1d). They then use this list to simulate a tryptic digest and construct a superset of plausible peptide sequences (1e). For a given spectrum, a small set of plausible candidates are constructed based on knowledge of the parent protein's mass and charge. Each of these plausible candidates are then scored against the mass spectrum (1g). After having scored all plausible candidates, the highest scoring peptide is referred to as the peptide-spectrum match (PSM). 

One of the most popular algorithms for scoring the agreement of a fragmentation spectra and a candidate peptide is the SEQUEST/COMET algorithm \cite{Eng:1994fj, Eng:2013ig}.  The algorithm explicitly enumerates the theoretically possible fragments for any given peptide and uses a modified cross-correlation to evaluate the extent of agreement between expected and observed mass-to-charge ratios of the fragments (1f). However, due to the complexity of the fragmentation process, these methods do not do a good job of taking into account the relative abundances of the different possible fragments. Based on a suite of factors known to affect ion formation, we expect that the structure of a peptide will greatly impact the relative abundances of different fragments~\cite{YingyingHuang:2004cl, Gucinski:2010kr}. A motivation for our deep learning architecture is to allow the model to encode information which is underutilized by conventional algorithms. 

The peptide which best matches a fragmentation spectra may nevertheless be incorrect, especially when the score of the match is poor. In practice it is often desirable to estimate the rate at which PSMs of a given score are false-positives. This is called the false discovery rate (FDR = $\mathbb{E}[\frac{FP}{FP + TP}]$) \cite{Storey:2003cj} and it can be estimated in the following way~\cite{Elias:2007dp, Kall:2008is}. First, a large set of ``dummy'' candidate peptides are scored against the spectra in the same process as outlined above. These dummy peptides are generated by reversing the genome and then simulating the same tryptic digest as is applied to the forward genome. These dummy peptides will be statistically similar to the original set, but we can say with high confidence that they will not be present in nature. Each fragmentation spectra is scored against all viable real and dummy peptides and the top-ranked PSM is saved.

At low scores where there is little overlap between the observed and theoretical spectrum, we expect the rate at which dummy peptides are identified as a PSM to approach their relative abundance in the set of candidates. Typically, researchers would like to identify a score cutoff such that scores higher than the cutoff have a FDR of less than 1\%. This can be found by the ratio of dummy to all peptides in the right-tail of the score distribution. In practice, a second round of re-ranking algorithms such as Percolator \cite{Kall:2007ea} are used to aggregate results across across many PSMs by using a linear-classifier to distinguish real and dummy peptides based on COMET scores and additional features.

\subsection{Weak Supervision For Peptide-Spectrum Matching}
We now discuss weak supervision in the context of the peptide-spectrum matching problem. We will first introduce a proxy task that we use to train our model, then we will discuss how we evaluate our model's performance. The strong ability of deep learning models to memorize data means that there are a number of potential pitfalls especially around traditional evaluation methods for these models.

\subsubsection{The Proxy Task}

We consider a proxy task that is intimately related to the problem of peptide-spectra matching. As discussed above, an existing search algorithm (in this case COMET) is applied to a dataset of spectra to attain, for each spectrum, a set of cross-correlation scores for each of the candidate peptides considered \cite{Eng:2013ig}. We use Percolator to rescore the data. As in the general case, this gives us a score for each peptide-spectrum match and the probability of a true match increases monotonically with the score. In this case, we may additionally explicitly estimate an error probability for each PSM\cite{Kall:2007ea, Kall:2007in}. We then construct a training set from the top-X (in this case $X = 3$) scoring peptides for spectra with an error probability $q \leq q^*$ (in this case $q^* = 50\%$). For each spectrum we assign the top-1 peptide a label $t = 1$ and the rest of the peptides a label $t=0$. The 2$^{\text{nd}}$ and 3$^{\text{rd}}$ top ranked matches are used as negative examples.

There is some subtlety when defining negative examples since it is easy to select negatives that are too easy for DeepMatch to distinguish. By selecting the candidates that caused as much confusion as possible for COMET, we know that the theoretical fragmentations for our positive and negative examples correlate with the measured spectrum. While ours is not a unique choice, it is one that we have found works well in practice.


\subsubsection{Model Selection and Evaluation}


To evaluate our model and perform hyperparameter selection we use the previously discussed competitive real-decoy peptide searches \cite{Elias:2007dp, Kall:2008is}. Thus, unlike in the case of training, we evaluate our model by scoring all the fragmentation events in an entire dataset. For each spectrum, we consider an identical set of candidate peptides to that assembled by COMET. We use our model to compute a score for each candidate paired separately with the spectrum. We then report the PSM for each fragmentation spectra. To summarize the error in our model, we compute the number of PSMs at 1\% FDR as described in section 4.1. 


While the decoy method is on solid footing when applied to classical peptide-spectrum matching algorithms,
for many choices of hyperparameters deep learning models will strongly tend to memorize peptides that it has seen. When such models are applied to the decoy method they perform exceptionally well, since they will preferentially select peptides that they have seen during training (which are all, by construction, real peptides). To get around this, we split our training and testing sets into different folds such that each fold contains spectra that were matched with a different disjoint subset of the total set of peptides. We then train our model on a subset of the folds and compute the number of peptides at 1\% FDR on the remaining folds. When we perform our evaluation, we use a culled set of candidates that excludes any peptides that may have been seen during training. In this case, models that memorize peptides cannot subvert the decoy method and so this is a well defined metric that can be used in hyperparameter tuning. 

To ensure that our models are not overfitting, we compare the number of peptides identified at 1\% FDR using the full set of candidates compared to the culled set of candidates. If these two quantities are similar then our model does not manipulate the metric significantly and we can proceed with confidence. Another test for overfitting
is to check that at low scores the false-discovery rate approaches $N_{\text{real}}/ N$. Since at low scores there is very little useful information from the spectra, DeepMatch should identify decoys at the rate given by their fraction of the overall candidates. 



\section{Experiments}

We take the best models after hyperparameter tuning and train them on data collected from several different species~\footnote{We intend to opensource the model and hyperparameters. More details are in the supplementary information.}. For each species we assemble a list of candidate peptides by direct enumeration. The size of this candidate list depends on the species in question but generally varies between $10^9$ and $10^{10}$ distinct peptide sequences whose length is between 5 and 40. We use 3800 bins to represent the mass spectrum. We then evaluate the model's performance on unseen datasets and compare the results to COMET. For a brief description of the datasets used see the supplementary information. Our hyperparameter search identified the best DeepMatch model to have a single layer bidirectional LSTM with hidden dimension 600 and a spectral embedding of dimension 60.

We compare the distribution of scores found by our network to the distribution identified by COMET.
\begin{figure}[!h]
\centering
\includegraphics[width=0.6\textwidth]{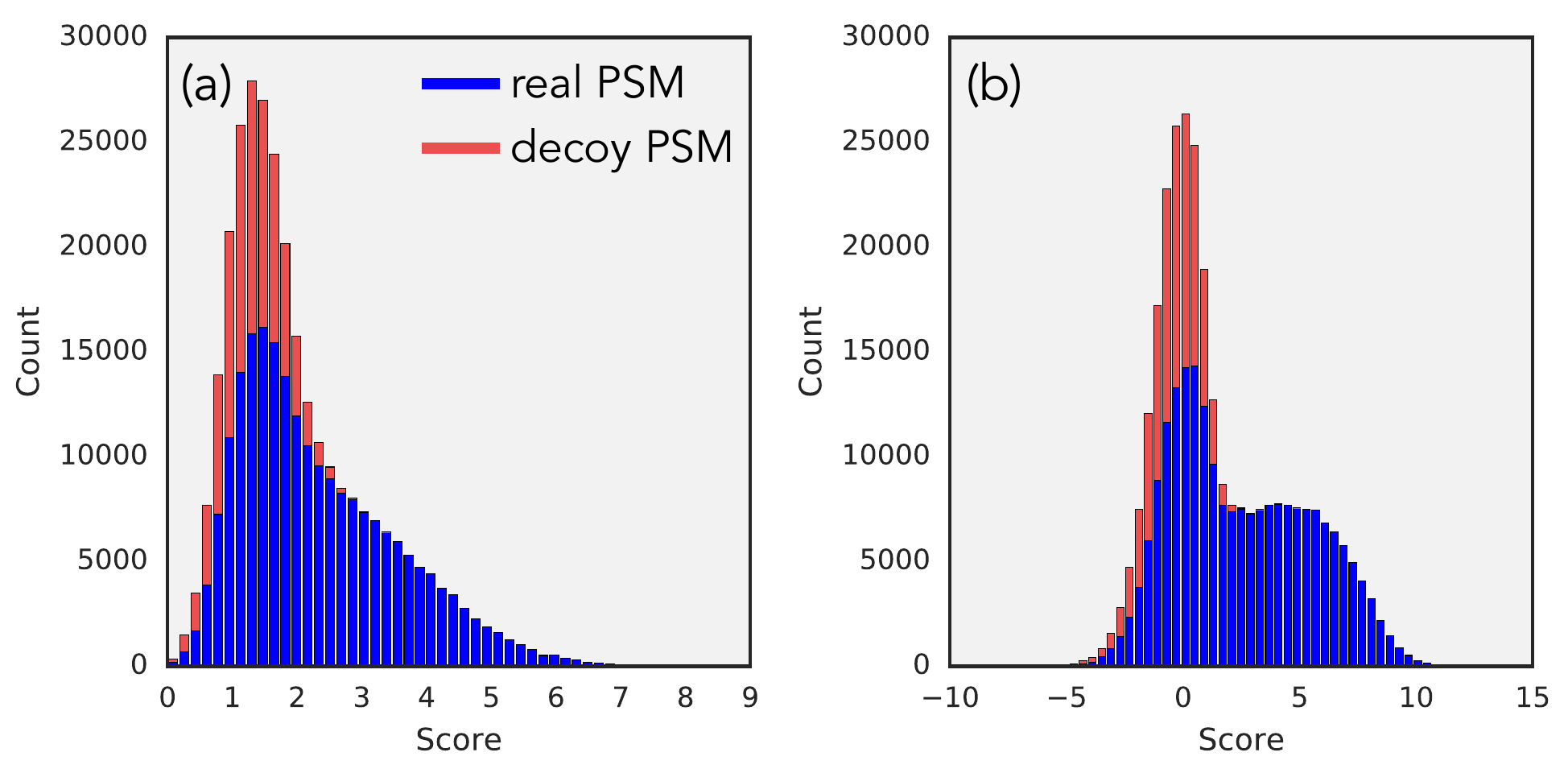}
\caption{\textbf{The distribution of scores for peptide-spectra matches.} These scores are colored by whether or not they were real peptide-spectra matches or decoys for (a) COMET (b) DeepMatch.}
\label{fig:figure_2}
\end{figure}
These distributions are shown in Figure~\ref{fig:figure_2}. We notice that by eye the distribution of scores identified by DeepMatch appears significantly more bimodal than the scores identified by COMET. We offer some analysis showing that our network's scores are well behaved in the supplementary information.

DeepMatch is significantly more successful at correctly identifying peptide-spectrum matches. 
\begin{figure}
\centering
\includegraphics[width=0.9\textwidth]{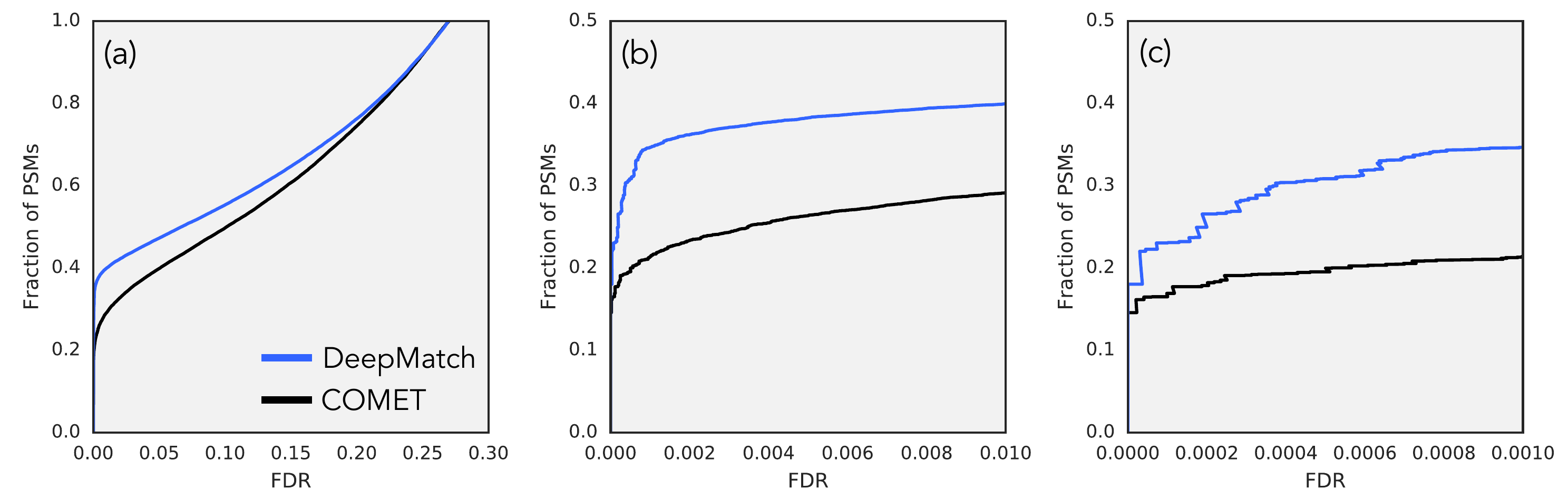}
\caption{\textbf{A comparison between DeepMatch and COMET using the FDR metric.} The different figures show the same plot (a) for the full range of FDRs, (b) for FDRs up to 1\%, and (c) for FDRs up to 0.1\%.}
\label{fig:figure_4}
\end{figure}
We observe in Figure~\ref{fig:figure_4} the number of peptide-spectrum matches at different FDRs. We observe that DeepMatch's performance dominates COMET across FDRs. Moreover, zooming in to very low false-discovery rates we see that DeepMatch monotonically outperforms COMET which implies that DeepMatch is well behaved at high scores. At 1\% FDR, we observe a 42\% improvement in the number of spectra matched to peptides.
\begin{table}[!h]
    \centering
    \begin{tabular}{ccc}\toprule
        \multirow{2}{*}{Search Algorithm}& \multicolumn{2}{c}{Fraction Matched At 1\% FDR} \\\cmidrule{2-3}
        & \hspace{2pc}Mouse\hspace{2pc} & \hspace{2pc}Yeast\hspace{2pc} \\\midrule
        COMET    & 29\% &  36\%  \\\midrule
        DeepMatch &  \textbf{40\%}    &   \textbf{47\%}   \\
        DeepMatch (No LSTM) &  39\%    &  45\%  \\
        DeepMatch (Shallow Readout) &  34\%    &  43\%  \\
        \bottomrule
    \end{tabular}
    \vspace{1pc}
    \caption{A summary of our model's performance as well as some baselines.}
    \label{tab:results}
\end{table}

The results above show that DeepMatch significantly outperforms COMET as measured by the number of PSMs at 1\% FDR as measured by the decoy search method. However, as we discussed above, the decoy search method gives an estimate of the FDR that can easily be manipulated if care is not taken. To confirm these results, we used DeepMatch to find PSMs on the ProteomeTools dataset~\cite{Zolg:2017jo}. ProteomeTools is a large synthetic dataset where ground truth labels are known and so the FDR may be computed directly.
\begin{table}[h]
\centering
    \begin{tabular}{ccccc}\toprule
    	Search Algorithm & Estimated FDR & Correct & Incorrect & True FDR\\\midrule
	DeepMatch & 1\% & 481638 & 5449 & 1.1\%\\ 
	COMET & 1\% & 374371 & 4633 & 1.5\%\\
        \bottomrule
    \end{tabular}
        \vspace{1pc}
    \caption{A comparison of the FDR as measured using the decoy search method and measured against ground truth labels.}
    \label{lab:proteometools}
\end{table}

Table~\ref{lab:proteometools} shows the results of this comparison. To do this, we score a number of 50 different samples from ProteomeTools using DeepMatch and COMET and we enumerate the PSMs identified at 1\% FDR as measured by the decoy search method. Among this subset of peptides, we then compute the FDR using the ground truth labels. We observe strong agreement between the estimated FDR using the decoy search method and the true FDR. 

\begin{figure}
\centering
\includegraphics[width=0.9\textwidth]{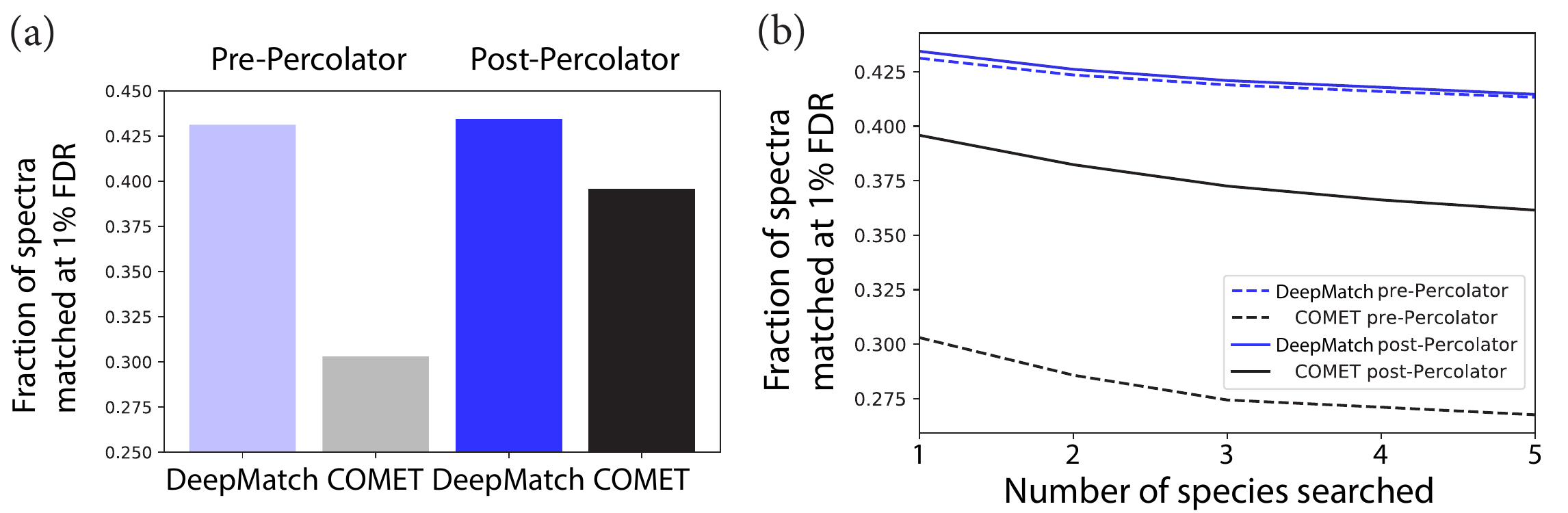}
\caption{\textbf{Performance of DeepMatch and COMET pre- and post-Percolator.} (a) Fraction of spectra that are matched to a peptide at a 1\% FDR when using COMET or DeepMatch for peptide-spectrum matching and optional recalibration using Percolator. (b) When searching spectra against an inflated list of target peptides (derived from additional species) the performance of DeepMatch models decay slower than COMET with or without Percolator.}
\label{fig:five_species}
\end{figure}

Percolator greatly improves the performance of COMET and barely influences the performance of DeepMatch. Nevertheless the performance of DeepMatch over COMET with Percolator applied is still substantial. The number of spectra that can be identified at a 1\% FDR is increased by 10\% when using DeepMatch over a combination of COMET and Percolator (Figure~\ref{fig:five_species}a). 
In many biological applications, the set of candidate peptides is significantly larger than the setting considered here. To test our model's performance in this regime we combined the candidate peptides from a number of species together. Since these additional candidate peptides will rarely be found in samples from mice, searching additional peptides should hurt our ability to detect true PSMs. However, we see little decay in the ability to recover matching with DeepMatch but a larger drop in performance when using COMET with or without Percolator. When searching five species simultaneously the pre-Percolator improvement of DeepMatch over COMET jumps to 54\% and the improvement of our network over COMET and Percolator rises to 15\%.

\section{Conclusion}

In this work we have introduced a method for training general machine learning models to identify peptides corresponding to a given mass spectrum based on noisy labels from classical algorithms. This can be used to study a class of inference problems in which an unknown sequence generates a distribution. We have additionally included a general discussion of a technique that can be used to leverage noisy labels to train our models to higher accuracy than the labels they were trained on. We observe strong performance on the important scientific problem of peptide-spectra matching and we believe that these techniques will be of broad use across a range of applications in the natural sciences.

\section*{Supplementary Material}

\subsection{Experimental Proteomics}

Biological studies often require analyses of samples containing over 10,000 proteins. The prevailing technology for summarizing the broad-scale biological state -- what proteins are present, if and how they are modified with post-translational modifications (PTMs), and how they vary across conditions -- is Liquid chromatography-mass spectrometry (LC-MS) based proteomics. To understand how this feat is achieved, we will briefly describe the experimental generative process of an LC-MS proteomics experiment before tying this to the inverse problem that mass spectrometry search engines solve.

Proteins are linear chains of 100-1000s amino acids. There are 20 of these amino acids which are shared by nearly all organisms. Each gene forms one-or-more proteins and the sequence of nucleotides in a gene determines the sequence of amino acids in its protein(s). Some amino acids in an intact protein may be further processed (e.g., through the addition of a phosphate or methyl group) altering both the size and function of the modified amino acid.

Intact proteins are too large and complex to be analyzed by mass spectrometry, so most LC-MS proteomics experiments first digest intact proteins into 10-100s of short chemically-similar peptides composed of typically between 8 and 30 amino acids. This digestion utilizes an enzyme, such as Trypsin, which cuts proteins at specific positions, in order to break the protein into a small number of predictable fragments. Having digested 10,000+ proteins into $\sim$100 peptides, there will now be 
$\sim$1,000,000 distinct peptides present in a sample, which researchers want to separately detect and quantify.

The goal of an LC-MS proteomics experiment is to first chemically separate distinct peptides and then to physically separate, isolate and fragment each peptide for the purpose of identification and quantification. Distinct peptides are first separated based on their chemical properties primarily through liquid chromatography. In liquid chromatography, peptides are passed through a chromatography column and the time that it takes the peptide to reach the end of the column will be determined by specific chemical properties of the peptide. Thus, at each point in time, a subset of the entire peptide pool will elute from the column and this subset of peptides will be further separated using mass spectrometry. In the mass spectrometer, peptides will be ionized and then precisely distinguished based on their mass-to-charge ($m/z$) ratio. This generates a spectra which pairs a set of $m/z$ ions with their associated ion current (abundance). These resolved peptides are finally isolated and then fragmented to generate a fragmentation spectra. A LC-MS proteomics experiments will generate >10,000 of such fragmentation spectra, and one of the chief challenges in computationally processing the resulting data is the "peptide-spectrum matching problem": attributing a peptide identity to each fragmentation spectra. 


The traditional method for inferring what proteins were present in a sample, from the collection of mass spectra is to use a search engine such as SEQUEST\citep{Eng:1994fj}, X!Tandem\citep{fenyo2003method} or Mascot\citep{perkins1999probability}. These search engines translate a database of genomic sequences from organisms being studied into protein sequences and digest them \textit{in-silico} using known digestion rules, to create an index of peptides. A short set of candidate peptide matches is generated for each mass spectrum from the peptide index using the $m/z$ values of the parent ions that were chosen for fragmentation and the mass of the peptides (using different possible integer charge values -- usually 1-4, or determined by interpreting the charge of the parent ion, through the mass spectrum itself).  Each candidate match is scored by generating a theoretical fragmentation spectrum and matching it to the observed one. Different search engines differ in how the theoretical fragmentation pattern is generated, and how the scoring functions are computed. However, for the most part, these fragmentation patterns used have been relatively simplistic -- assuming all theoretical peaks are observed, and with equal intensity -- when, in fact, peptide fragmentation is quite probabilistic, and depends on a variety of factors, such as the sequence of the peptide. This is the main limitation we tackle in this paper using the available datasets to learn a good scoring function and a predictive model for the fragmentation patterns.

Once candidate peptides matches are generated, they can be further filtered out by using manual or automated rules. Manual rules often use criterion based on numbers of peptides per proteins, while automated rules can use algorithms such as Protein Prophet \citep{keller2002empirical} or Percolator\citep{Kall:2007ea} to further filter results from the first pass search. Percolator uses a support vector machine (SVM) to automatically train a second model on search results from the first pass search on a database with real peptides and decoy peptides (generated from a fake database -- usually, a reversed protein dataset). The SVM is trained to separate high scoring matches from the forward direction sequence (label = 1) from the high scoring matches from the reverse direction (label = 0). The features used by Percolator as generally high level features of the matches, such as $m/z$ difference between parent ion and theoretical $m/z$. The trained SVM is then used to clean up results from the first pass. Such a strategy can of course be used to clean up results from our methods as a second pass.

\subsection{Model Components}

We will now discuss a number of different components that we considered within our overall model family. Note that each of the four pieces above can be changed separately while adhering to the same overall structure as long as the dimension of the input and output are compatible.

\subsubsection{The Fragment Representation}

We propose two different methods of converting the amino acid representation into a fragment representation.
\begin{enumerate}
\item \underline{Trivial Representation}:

As a baseline, we propose a trivial embedding where the amino acid representation is converted into a fragment representation via a projection. To do this we take $F:\mathbb R^{L\times A}\to\mathbb R^{(L+1)\times 2A}$ defined by,
\begin{equation}
F(x_1,\cdots x_L) = \left(\begin{matrix}x_1 & \cdots & x_L & 0\\ 0 & x_1  & \cdots & x_L \end{matrix}\right)
\end{equation}
Thus, the fragment $f_i$ corresponding to a breakage between amino acid $i-1$ and $i$ is given by the concatenation of the amino acid representation before and after the breakage.

\item \underline{LSTM Representation}:

In principle, we would like the fragment representation to be able to model the fragmentation probabilities between different amino acids. However, these probabilities depend not only on the pair of amino acids on either side of the breakage point, but on the structure of the entire peptide. To this end, we propose using a deep bidirectional LSTM with $K$ layers and hidden dimension $F/2$ to construct the fragment representation that can take into account this nonlocal information. In this case the fragment representation is defined by,
\begin{align}
(z_1,\cdots, z_L) &= \text{Bi-LSTM}(x_1,\cdots, x_L)\\
F(x_1,\cdots x_L) &= \left(\begin{matrix}z_1 & \cdots & z_L & 0\\ 0 & z_1  & \cdots & z_L \end{matrix}\right).
\end{align} 
\end{enumerate}

\subsubsection{The Spectral Representation}

We propose two different methods of converting from the fragment representation to the spectral representation. In one case, we use theoretical knowledge about the possible ways that the peptide can fragment. The other  approach is agnostic to the underlying physical process. While we see significantly stronger performance when we use theory to constrain the form of our model, the agnostic approach may be more suitable when an explicit fragmentation cannot be enumerated. We also note that there is some overhead computational cost in precomputing the possible fragments.

\begin{enumerate}
\item \underline{Agnostic Representation}:

Here we use a fully-connected network $g:\mathbb R^F\to\mathbb R^M$ to map each of the fragment representation into a corresponding contribution to the spectral representation. Thus, we write
\begin{equation}
Y(f_1,\cdots,f_{L+1}) = \sum_\ell g(f_\ell).
\end{equation}
This representation has $H = 1$ and potentially offers more interpretability since it can be normalized and interpreted as a probability distribution. Unfortunately, we were unable to get good performance using an agnostic representation that did not take explicit fragmentation patterns into account at the time of writing.

\item \underline{Ion Representation}:

In this case we wish to take into account the fact that physically there are a finite number of ways in which a peptide may fragment. For each peptide we therefore enumerate all possible ways that the fragmentation can occur along with the theoretical mass-to-charge ratio that the resulting fragment will have. 

To construct this enumeration, we first note that fragmentation occurs via the breaking of a bond between two amino acids. Moreover, between each pair of amino acids there are a number of different ways in which the fragmentation can occur (with different associated theoretical masses). In our experiments we use the three most likely breakages yielding so-called``b'', ``y'', and ``a'' ions. Additionally fragments can have different charges associated with them. In our case we consider charges $Z=1,2$. Finally, can also have different neutral losses associated with them (where a molecule disociates from the peptide during fragmentation). In our case we consider fragments that have lost an $\text{NH}_3$ molecule or an $\text{H}_2\text{O}$ molecule.

Together this gives a combinatorial number of different ways that each breakage may occur. We enumerate these different possibilities with integers $0\leq c \leq C\equiv 18$.  For each possible breakage we construct a fully connected neural network $g_c:\mathbb R^F\to\mathbb R^H$ for some hidden dimension $H$. Note that since fragmentation may occur between any pair of amino acids, there are a total of $C(L-1)$ ways that the peptide can be fragmented overall. We let $m_{ic}$ be the mass/charge ratio of the fragment of type $c$ that occurs between amino acid $i$ and $i+1$. We then define the ion representation by $G\in\mathbb R^{M\times H}$, where,
\begin{equation}
G_k = \sum_{\substack{(i, c)\\\text{ s.t. }\\ m_{ic}\in[M_k, M_{k+1}]}}g_c(f_i).
\end{equation}
Note that since the total number of possible fragments is small compared to the number of $m/z$ bins, $G$ will typically be very sparse.
\end{enumerate}

\subsubsection{Readout}
We consider two different readout networks.
\begin{enumerate}
\item \underline{Trivial Readout}:

As a baseline, we consider a trivial readout where a fully-connected network is applied directly to the concatenated spectral representation. In this case the score is given by, $s = w(G)$ for some fully connected $w:\mathbb R^{M\times H}\to\mathbb R$.

\item \underline{VGG Readout}:

In general, we would like our network to learn complex correlations between the empirical spectrum and the spectral representation. To that end we use a copy of the VGG network~\cite{simonyan2014very} that has been successful in computer vision models. Thus, we take $s = \text{VGG}(G)$.
\end{enumerate}

\subsection{Dataset Description}

We consider data from a number of species taken from the ProteomeXchange database~\cite{vizcaino2014proteomexchange} containing raw spectra. In particular we considered the datasets PXD005953 (Mouse), PXD002801 (Mouse), PXD001334 (Yeast), PXD002875 (Yeast), PXD003033 (Yeast), PXD005253 (Human), and PXD003389 (Human). These datasets varied significantly in size with the smallest having around 30K scans and the largest having $\sim10$M scans. For each species we assemble a list of candidate peptides by direct enumeration. The size of this candidate list depends on the species in question but generally varies between $10^9$ and $10^{10}$ distinct peptide sequences whose length is between 5 and 40. We discretize the mass spectra using bins between 100 Daltons and 2000 Daltons of width 0.5 Daltons. This gives us 3800 bins in our discretized mass spectrum.

\subsection{Network Diagnostics}

We show that both the network scores and the COMET scores display similar characteristics.
\begin{figure}[!h]
\centering
\includegraphics[width=0.7\textwidth]{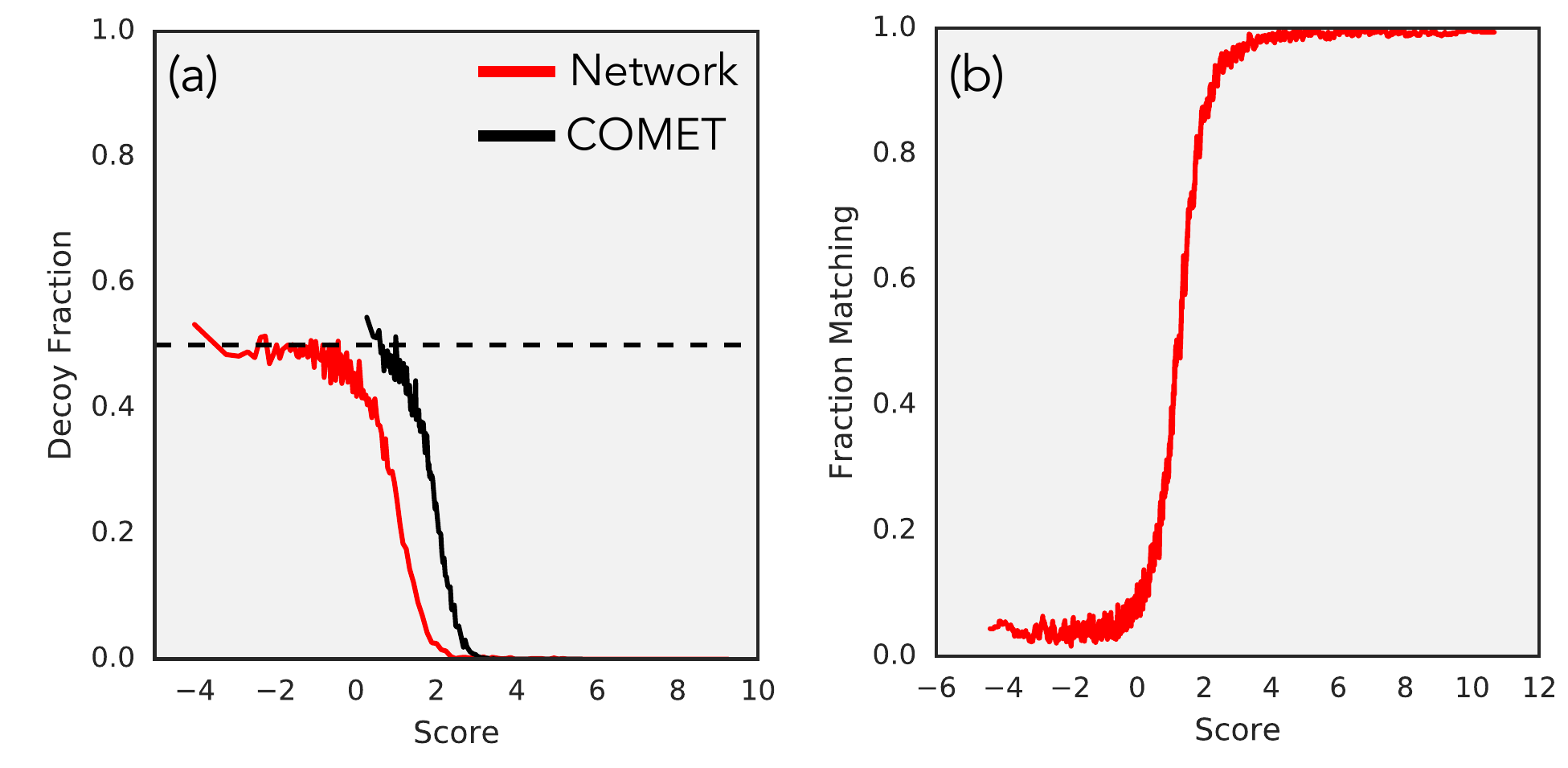}
\caption{\textbf{Characteristics of the network scores.} (a) The fraction of decoys identified as top-1 peptide-spectra matches by the network and COMET. We see that at low scores both approach $\frac{1}{2}$ which is shown by a dashed line. (b) The fraction of peptide-spectrum matches that have agreement between the network at COMET. We see that at high scores the two methods almost always agree.}
\label{fig:figure_3}
\end{figure}
To this end we show, in fig.~\ref{fig:figure_3} (a), the decoy fraction of top-1 peptide-spectrum matches as a function of the score assigned to the match. We see that we have successfully avoided selecting networks that memorize peptide data since our network and COMET both approach a decoy fraction of $\frac{1}{2}$ at low scores. Next, we calculate, as a function of the score, the fraction of peptide-spectrum matches where the network gave the same prediction as COMET. This is shown in fig.~\ref{fig:figure_3} (b). We see that at high scores the two assignments almost always agree. At low scores, the rate at which the two agree approaches the random value.

\small
\bibliographystyle{plainnat}
\bibliography{Bibs/sean_partial_bib}

\end{document}